\documentclass[%
 reprint,
 amsmath,amssymb,
 aps,
]{revtex4-2}

\usepackage{dcolumn}
\usepackage{bm}
\usepackage{epsfig}
\usepackage{xcolor}

\begin{document}

\preprint{APS/123-QED}

\title{Evolution of honesty in higher-order social networks}

\author{Aanjaneya Kumar}
\email{kumar.aanjaneya@students.iiserpune.ac.in}
\affiliation{Department of Physics, Indian Institute of Science Education and Research, Dr. Homi Bhabha Road, Pune 411008, India}

\author{Sandeep Chowdhary}
\email{chowdhary_sandeep@phd.ceu.edu}
\affiliation{Department of Network and Data Science, Central European University, 1100 Vienna, Austria}

\author{Valerio Capraro}
\email{v.capraro@mdx.ac.uk}
\affiliation{Department of Economics, Middlesex University, The Burroughs, London NW4 4BT, United Kingdom}

\author{Matja{\v z} Perc}
\email{matjaz.perc@gmail.com}
\affiliation{Faculty of Natural Sciences and Mathematics, University of Maribor, Koro{\v s}ka cesta 160, 2000 Maribor, Slovenia}
\affiliation{Department of Medical Research, China Medical University Hospital, China Medical University, Taichung, Taiwan}
\affiliation{Alma Mater Europaea ECM, Slovenska ulica 17, 2000 Maribor, Slovenia}
\affiliation{Complexity Science Hub Vienna, Josefst{\"a}dterstra{\ss}e 39, 1080 Vienna, Austria}

\date{\today}

\begin{abstract}
Sender-receiver games are simple models of information transmission that provide a formalism to study the evolution of honest signaling and deception between a sender and a receiver. In many practical scenarios, lies often affect groups of receivers, which inevitably entangles the payoffs of individuals to the payoffs of other agents in their group, and this makes the formalism of pairwise sender-receiver games inapt for where it might be useful the most. We therefore introduce group interactions among receivers, and study how their interconnectedness in higher-order social networks affects the evolution of lying. We observe a number of counterintuitive results that are rooted in the complexity of the underlying evolutionary dynamics, which has thus far remained hidden in the realm of pairwise interactions. We find conditions for honesty to persist even when there is a temptation to lie, and we observe the prevalence of moral strategy profiles even when lies favor the receiver at a cost to the sender. We confirm the robustness of our results by further performing simulations on hypergraphs created from real-world data using the SocioPatterns database. Altogether, our results provide persuasive evidence that moral behaviour may evolve on higher-order social networks, at least as long as individuals interact in groups that are small compared to the size of the network.
\end{abstract}

\maketitle


\section*{Introduction}
Flow of information from a source to its destination is ubiquitous, and fundamental to life at all levels -- from signalling at the cellular level for coordinating the activities of cells to communication between people in and across societies. Particularly crucial is communication and transmission of honest signals between different parts of a system for maintaining its efficient functioning as epitomized by colonies of ants and honey bees and also by the \emph{super-cooperating} humans \cite{nowak_supercooperators_2011}. However, the breakdown of communication and dishonest signalling can severely hamper the operations of a system and lead to undesirable consequences. A striking example of this is how misinformation and rumours can propagate fear and paranoia during times of crises, like the coronavirus disease 2019 (COVID-19) pandemic, and amplify the hardship that come along with it. Needless to say that tackling the problem of misinformation and fake news stands as one of the most important challenges of our time.

While the study of fake news has attracted a lot of attention over the last years \cite{pennycook2019fighting,pennycook2019lazy,pennycook2020fighting}, the literature on signalling and deception has a much longer history \cite{skyrms_signals}. Evolutionary biologists have developed several models to study diverse scenarios such as predator–prey signaling \cite{chivers1998chemical,leal1997signalling,ramesh2018evolution}, sexual signaling \cite{rooker2018evolution,sun2020costly}, and the interactions between siblings \cite{caro2016sibling} and between parents and offsprings \cite{dugas2016baby}. On the other hand, the focus of economists and psychologists has been towards the development of tasks which would allow us to quantify (dis)honesty. Some examples include the die rolling paradigm \cite{fischbacher2013lies}, the matrix search task \cite{mazar2008dishonesty}, the Philip Sidney game \cite{smith1991honest}, and the sender-receiver game \cite{gneezy2005deception}. The latter game, in particular, has received a great deal of  attention in the last years.

The sender-receiver game is a classic example of a game with asymmetric information, and provides us with a paradigm to explore strategic interactions between two types of agents: \emph{senders}, who possess a piece of information and can either honestly or dishonestly communicate it to the second type of agents; \emph{receivers}, who can choose to either believe or not believe the message sent to them. A particularly interesting feature of this game is that it allows us to distinguish among different types of lies, based on whether lying has positive or negative consequences for the players \cite{erat2012white}. Not surprisingly, it has attracted a lot of research attention from behavioural scientists \cite{gneezy2005deception,sanchez2007experimental,dreber2008gender,peeters2008rewards,cohen2009groups,rode2010truth,erat2012white,cappelen2013we,gunia2012contemplation,gneezy2013measuring,kouchaki2013seeing,sheremeta2013liars,levine2014liars,biziou2015does,greenberg2015promoting,levine2015prosocial,roeser2016dark,capraro2017does,capraro2018gender,gneezy2018lying,lohse2018deception,capraro2019time,speer2020cognitive,abe2020overriding} and, more recently, from physicists. Recent studies \cite{capraro2019evolution,capraro2020lying} using Monte Carlo simulations following the replicator dynamics have provided an extensive theoretical study of the sender-receiver game in well-mixed and structured populations to complement the massive amount of experimental data available.

Although important, these studies consider only one-to-one interactions, and a major assumption is that the payoffs of individual receivers in a population depends solely on their own strategy, and the strategy of the sender they interacted with. However, since people in a society are interconnected, it is inevitable that the payoffs of each receiver would not only depend on whether they were deceived, but also on whether the other receivers around them got deceived. In panels B and C of Fig.~1, we emphasize this point. This realization, though simple, has profound consequences for models of strategic interactions as the fitness of an individual in a population is entangled with the fitness of other people in the population. This calls for a new modelling framework, which goes beyond the simple consideration of only pairwise interactions between agents, and takes into account the effects of group interactions to provide an improved description of reality \cite{battiston_networks_2020}.  While this paradigm of including \emph{higher order interactions} has successfully been incorporated in the study of the evolution of cooperation in the form of the Public Goods Game \cite{alvarez2021evolutionary}, there is a strong need for a systematic study of such group effects in other problems of strategic interactions, such as signalling games, where higher-order interactions are inevitable, and must be accounted for. This encourages us to explore the role of group interactions in the evolution of honesty and lying in the sender-receiver game. We note that in several other contexts, higher-order interactions have already been shown to lead to novel collective phenomena as in the case of synchronization phenomena~\cite{bick2016chaos,skardal2019abrupt, millan2020explosive, lucas2020multiorder}, random walks ~\cite{carletti2020random,schaub2020random}, consensus~\cite{neuhauser2020multibody, iacopini2021vanishing}, and ecological dynamics~\cite{bairey2016high, grilli2017higher}.

The plan of our paper is as follows: In Section \ref{se:model}, we extend the sender-receiver game to incorporate group effects. In Section \ref{suse:well-mixed}, we study the evolution of different strategies in a well-mixed population and explore the parameter space of the game to unravel rich and novel results. In Section \ref{suse:hypergraphs}, we explore the evolution of strategies on a class of hypergraphs, namely the hyperrings. The hyperrings also allows us to investigate the role played by the size of the hyperedges on the evolution of the strategies. In Section \ref{suse:real world} we examine the evolution on four `real world' hypergraphs, built using the SocioPatterns dataset. We conclude with a discussion and a brief outlook for future research in Section IV.

\begin{figure}
    \centering
    \includegraphics[scale=0.175]{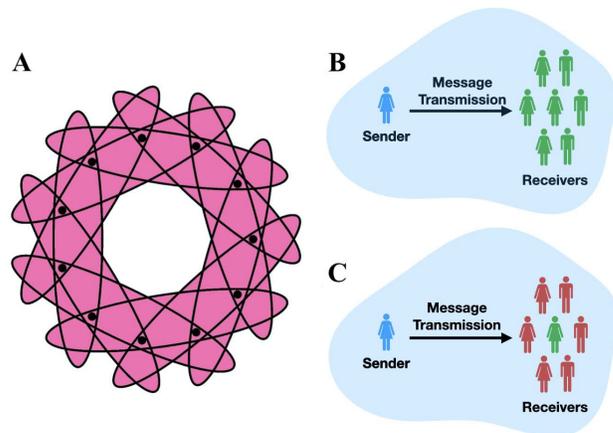}
    \caption{Schematic of different aspects of our work. Panel A: the hypergraph considered for our simulations -- the hyperring. The one shown in the figure consists of 11 nodes with each hyperedge containing 4 nodes. Panels B and C denote two possible scenarios which exemplify the need to incorporate group interactions -- suppose a sender sends a deceitful message to a group of receivers, which could possibly lead them to disregard the social distancing norms in the pandemic. The situation described in panel B depicts a scenario where none of the receivers were deceived (all marked in green) whereas panel C represents a scenario where a receiver who was not deceived (centre, marked in green) is surrounded by receivers who were deceived. It is clear that the central receiver in panel C should obtain a lower payoff than the one in panel B as the pandemic is being facilitated (by the deceived receivers) and restrictions will be extended for everyone. This \emph{group effect} is not captured by the pairwise models considered in literature. }
\end{figure}

\section{Mathematical model}\label{se:model}

\subsection{The Sender-Receiver game}\label{suse: sender receiver game}

We first define the canonical $2$-player sender-receiver game, as introduced by Erat and Gneezy \cite{erat2012white}. In this game, the sender first rolls a die and observes the outcome of the roll which can be any of the six possible outcomes in $\{1,2,3,4,5,6\}$. The sender then sends a message about the outcome, which can be the truth (T) or a lie (L), and the message is communicated to the receiver. After receiving the message, the receiver chooses a number between $1$ and $6$. If this number is equal to the actual outcome of the die, without loss of generality, we set the payoffs of the sender and receiver, both, to be 0. In case this number is different from the actual outcome of the die, then the sender gets a payoff $s$ and the receiver gets $r$. This game can be easily reduced to a game with two strategies for each player. Indeed, the payoff of the receiver essentially depends only on whether they choose to believe (B) the message sent by the sender, or not (N). We can rewrite the payoff bimatrix of this game as follows:

\begin{center}
\begin{tabular}{ |c|c|c| }
 \hline
  & B & N \\
 \hline
 ~T~ & ~~$0,0$~~ & ~~$s,r$~~ \\
 ~L~ & ~~ $s,r$~~ &~ ~$\frac{4}{5}s$ ,$\frac{4}{5}r$ ~~ \\
 \hline
\end{tabular}
\end{center}
where the ratio $\frac{4}{5}$ comes from the fact that, if the sender lies and the receiver does not believe the sender, then the receiver reports a wrong outcome of the die with probability $\frac{4}{5}$.

The fact that the sender-receiver game is essentially a game with two players, each of which has only two strategies, makes it suitable to study using methods of statistical physics, and the Monte Carlo method in particular. Another reason why the sender-receiver game is becoming increasingly popular is that it allows to study different types of lies, depending on the values of $s$ and $r$:
    \begin{enumerate}
    \item[] \emph{Pareto white lies} are those that benefit both the sender and the receiver: $r,s>0$.
    \item[] \emph{Altruistic white lies} are those that benefit the receiver at a cost to the sender: $r>0$, $s<0$.
    \item[] \emph{Black lies} are those that benefit the sender at a cost to the receiver: $r<0$, $s>0$.
    \item[] \emph{Spiteful lies} are those that harm both the sender and the receiver: $r,s < 0$.
   \end{enumerate}
The distinction among different types of lies is useful also for the equilibrium analysis, which indeed depends on the type of lie. In the domain of spiteful lies, there are two equilibria in pure strategies -- $(T,B)$ and $(L,N)$ -- and one equilibrium in mixed strategies -- $(\frac16 T + \frac{5}{6}L, \frac16 B + \frac{5}{6}N)$. In the domain of altruistic or black lies, there is only one equilibrium: $(\frac16 T + \frac{5}{6}L, \frac16 B + \frac{5}{6}N)$. In the domain of Pareto white lies, there are two equilibria in pure strategies -- $(T,N)$ and $(L,B)$ -- and one equilibrium in mixed strategies, once again, $(\frac16 T + \frac{5}{6}L, \frac16 B + \frac{5}{6}N)$. The cases $r = 0$ and/or $s = 0$ are straightforward, because the corresponding players are indifferent between the two available strategies.

\subsection{Introducing group interactions}\label{suse: group interactions}

We now extend the canonical sender-receiver game to allow for multiple receivers, who interact among themselves, meaning that the payoff of each receiver does not only depend on whether the receiver herself believes the message sent by the sender, but also on whether the other receivers believe this message. In particular, we assume that the payoff obtained by each receiver is the sum of two components: the payoff due to the individual `pairwise' interaction with the sender ($\Pi_I$), and the payoff due to the group interaction with the other receivers ($\Pi_G$).

The pairwise payoff, $\Pi_I$, is defined to be identical to the payoff of the standard sender-receiver game described in Section \ref{suse: sender receiver game}.


To account for the group payoff, $\Pi_G$, we introduce some terminology and some notation. We say that a receiver is \emph{deceived} by the sender if they report a wrong outcome of the die. We denote $x_W$ the fraction of receivers who are deceived by the sender (excluding the receiver for which we are calculating the payoff); where $W$ stands for \emph{wrong} as the receiver reports the wrong outcome of the die. When a receiver is not deceived by the sender, we define their group payoff to $\Pi_G= k_Rx_W$; where $k_R$ is a fixed real number, and $R$ stands for \emph{right}, because the receiver reports the right outcome of the die. When the receiver reports the wrong outcome of the die, we denote their group payoff as  $\Pi_G=k_Wx_W$, where $k_W$ is, again, a real constant.

To summarize, the total payoff, $\Pi_I + \Pi_G$, obtained by each receiver can be tabulated as:
   \begin{center}
\begin{tabular}{ |c|c|c| }
 \hline
  & B & N \\
 \hline
 ~T ~& ~~~$ 0+k_Rx_W$ ~~ & $ r+ k_W x_W$ \\
 ~L ~& ~~~$r+k_Wx_W$ ~~ & ~~ $ \frac{1}{5}k_Rx_W + \frac{4}{5}(k_Wx_W+r) $ ~~ \\
 \hline
\end{tabular}
\end{center}
Of course, as $r, k_R, k_W$ vary in $\mathbb R^3$, one can have different prototypical cases. A detailed discussion of the different types of group interactions captured by this model is presented in our discussion section. In principle, the model allows for various choices of group payoffs $\Pi_G$ received by individuals, which can be defined to be general functions $f_R(x_W)$ and $f_W(x_W)$, depending on whether the receiver was deceived by the sender or not. For simplicity, we make the choice of assuming the payoffs to be linear functions of the fraction of deceived receivers in the group ($f_R(x_W)=k_Rx_W$ and $f_W(x_W)=k_Wx_W$). Clearly, setting $k_R=k_W=0$, one reduces the game to the standard sender-receiver game.

So far, we have defined only the payoff of the receiver. We define the payoff of the sender to be simply the sum of the payoffs that they obtain in their interactions with each receiver.

\subsection{The Monte Carlo method for well-mixed populations}\label{suse:monte carlo well-mixed}

We consider a sender-receiver game among $n$ agents. Initially, each agent is randomly assigned one of the four pure strategy profiles $(T,B), (T,N), (L,B), (L,N)$. Then one agent is randomly selected to play as the sender. The $n$ agents then receive a payoff from the $n$-player sender-receiver game played with the selected agent in the role of sender and all the other agents in the role of receivers. At the end of this interaction, another agent is selected to play in the role of sender. We repeat this procedure $n$ times so that, at the end of this Monte Carlo step each agent has played the role of sender exactly once. This concludes one Monte Carlo step.

At the end of a Monte Carlo step, for each agent we randomly select another agent. Then the first agent copies the strategy of the second agent with a probability that depends on the difference between their accumulated payoffs in the last $n$ rounds. In particular, if player $P_1$ and its randomly selected pair, player $P_2$, collect payoffs $\Pi_{P_1}$ and $\Pi_{P_2}$, respectively, then $P_1$ copies the strategy of $P_2$ with probability
\begin{equation}
    w=\frac{1}{1+\exp\left[\left(\frac{\Pi_{P_1}}{n}-\frac{\Pi_{P_2}}{n}\right)/K\right]}
\end{equation}
where K quantifies the uncertainty/error in strategy adoptions. In real-world settings, one would expect that agents would try to copy their neighbours who are performing better than them. However this imitation would have its limitations and imperfections, and with a small probability, agents can also copy the strategy of poorly performing neighbours. In our simulations, we choose $K=0.1$, unless explicitly stated otherwise.

To reach sufficient accuracy, we simulate large system sizes, with $n = 500$, as well as long enough thermalization and sampling times, of $10^5$ Monte Carlo steps.

\subsection{The Monte Carlo method for hypergraphs}\label{suse:monte carlo hypergraphs}

To study the effect of spatial correlations on the evolution of strategies in the $n$-player sender-receiver game, we simulate the game on hypergraphs. A hypergraph is a generalization of a graph, where edges, instead of connecting two nodes, can connect any number of nodes. Formally, a hypergraph is a pair $H=(X,E)$, where $X$ is a non-empty set of nodes and $E$ is a family of non-empty subsets of $X$, called hyperedges. Nodes who belong to the same hyperedge are said to be neighbors.

The sender-receiver game is simulated on hypergraphs as follows.

We treat each hyperedge as a well-mixed group. Within one Monte Carlo step, we first select a hyperedge and all players in that hyperedge play the group sender-receiver game with each other such that each player in the hyperedge gets to play the role of a sender exactly once. To account for heterogeneties in sizes of the hyperedge, the payoff obtained by each agent in the hyperedge is divided by the number of agents in the hyperedge. This is repeated until we go over all hyperedges. Players accumulate payoffs from all hyperedges they are part of. This brings us to the imitation step, where we go over each of the $n$ players in the hypergraph one by one, and each player $P_1$ copies the strategy of a randomly chosen neighbour $P_2$ with probability
\begin{equation}
    w=\frac{1}{1+\exp\left[\left(\frac{\Pi_{P_1}}{\mathcal{K}_1}-\frac{\Pi_{P_2}}{\mathcal{K}_2}\right)/K\right]}
\end{equation}
where $\mathcal{K}_i$ denotes the number of hyperedges node $i$ is part of.

\section{Results}
\begin{figure*}
    \centering
    \includegraphics[scale=0.7]{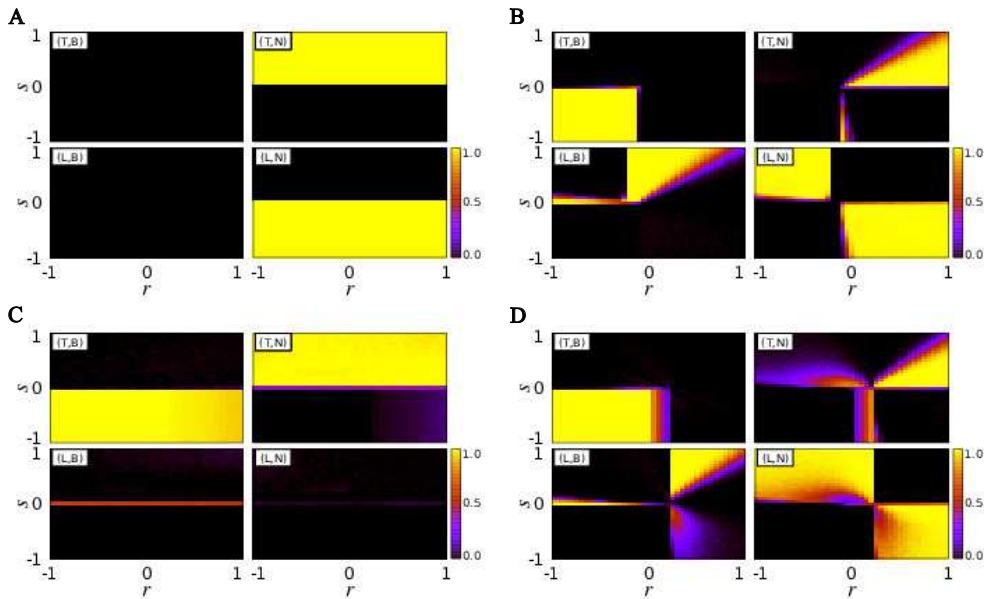}
    \caption{Final densities of the four strategy profiles, $(T,B)$, $(T,N)$, $(L,B)$ and $(L,N)$, in well-mixed populations for panel A: $k_R=0,k_W=10$, panel B: $k_R=0,k_W=0.25$, panel C: $k_R=10,k_W=0$ and panel D: $k_R=0.25,k_W=0$. We show that the presence of group interactions can lead to evolution which is quite different from the case where only pairwise interactions are considered (compare with Fig.~S1 in Ref.~\cite{SI}, and Fig. 2 of Ref. \cite{capraro2019evolution}).}
\end{figure*}
The results section is structured as follows. In Section \ref{suse:well-mixed}, we simulate the evolutionary group sender-receiver game on well-mixed populations, and we compare the results with those already obtained for the standard pairwise sender-receiver game \cite{capraro2019evolution}. We will see that, already in the well-mixed case, several new features emerge with the addition of group interactions. Then, in Section \ref{suse:hypergraphs} we study the evolution in the group sender-receiver game on a particular class of hypergraphs, hyperrings, and we will see that, compared to the well-mixed populations, further new features emerge. Finally, in Section \ref{suse:real world} we explore the evolution of the group sender-receiver game in four real world hypergraphs, and we show that the results qualitatively confirm those obtained in the hyperrings. The main result will be that the spatial structure leads to the evolution of the \emph{moral strategy} $(T,B)$, whereby the sender sends a truthful message and the receiver trusts the message sent by the sender, even for small values of $k_R$ and independently of $k_W$. Moreover, the moral strategy $(T,B)$ can more easily emerge in small groups than in large ones.

\subsection{Well-mixed populations}\label{suse:well-mixed}

We start by reporting the densities in the stationary state of the four strategy profiles (T,B), (T,N), (L,B), and (L,N), as a function of the parameters $s$, $r$, $k_R$ and $k_W$ in a well-mixed population consisting of $500$ agents.

First, we isolate the effect of $k_W$ by setting $k_R=0$. We report two simulations, one for $k_W=10$ and one for $k_W=0.25$. We choose these two parameter values as they provide key insight into the behaviour of the model, and how evolutionary dynamics is affected by incorporation of higher-order interactions. Since the absolute values of $s$ and $r$ (the pairwise payoffs) lie between $0$ and $1$, choosing the value of the group payoff strength to be $k_W=0.25$ allows us to probe the model behaviour in the case in which the payoffs associated with the group interactions are comparable to those associated with the pairwise interaction. Similarly, the value $k_W=10$ allows us to explore the behaviour of the model when group interactions bring a much higher payoff than pairwise interactions, and thus are the dominant factor in determining the evolution of the strategies.

When $k_W=10$, it is extremely beneficial for each receiver to be deceived, whenever at least another receiver is deceived. This suggests that the believing strategy $B$ should quickly vanish and, at the stationary state, receivers should never believe the message sent by the sender. This intuition is confirmed by the simulations (Fig. 2, panel A). The strategy played by senders can also be easily determined. It suffices to observe that, when $s>0$, senders would want to maximize their chances of deceiving the receivers and hence would choose to tell the truth ($T$) to a population of non-believers ($N$), making $(T,N)$ the strategy profile with maximal payoff. On the contrary, when $s<0$, senders have an incentive to lie ($L$) to a population of non-believers ($N$), making $(L,N)$ the stable strategy profile. Hence, for $s>0$, $(T,N)$ is the evolutionarily stable strategy, while otherwise it is $(L,N)$. We report this result in panel A of Fig. 2, where we plot the stationary frequencies of different strategies, for $s$ and $r$ taking values between $-1$ and $1$.
\begin{figure*}
    \centering
    \includegraphics[scale=0.7]{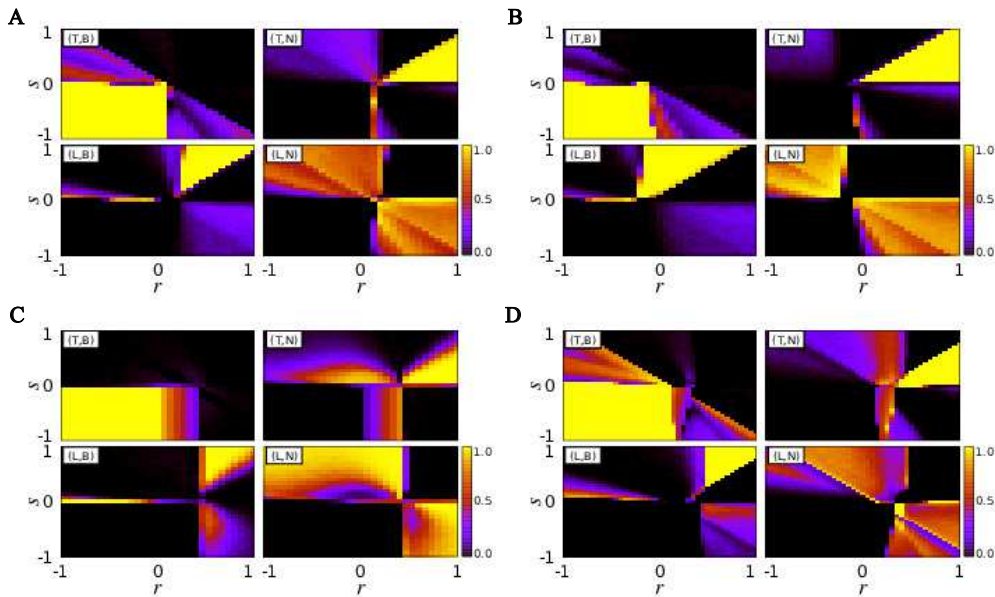}
    \caption{Final densities of the four strategy profiles, $(T,B)$, $(T,N)$, $(L,B)$ and $(L,N)$, in the hyperring where each hyperedge has size 4, for (panel A) $k_R=0.25,k_W=0$ and (panel B) $k_R=0,k_W=0.25$.  We also provide a comparison between $(k_R,k_W)$ = $(0.25,-0.25)$ for (panel C) well mixed and (panel D) hyperrings. Here we note that there is an emergence of the moral profile on the hyperring, which is not present in well mixed populations for $(s>0,r<0$ and $s<0,r>0)$. }
\end{figure*}

When $k_W=0.25$, it is still beneficial for each receiver to be deceived, whenever the other receivers are deceived. However, since this time both group interaction parameters, $k_R$ and $k_W$, are small, we observe a more nuanced evolution that looks similar to the one already reported for pairwise interactions (see Fig. S1 \cite{SI}, and also Fig. 2 in Ref~\cite{capraro2019evolution}), apart from a small shift. This result in reported in panel B of Fig. 2.
Next, we isolate the effect of $k_R$, by setting $k_W=0$. We report two simulations, one for $k_R=10$ and one for $k_R=0.25$.

When $k_R=10$, it is extremely beneficial for a receiver not to be deceived, whenever the other receivers are deceived. Starting from random initial conditions, it is easy to argue that believers have an initial advantage as they are less likely to be deceived. If, in addition, $s<0$, then the `moral strategy profile' $(T,B)$, whereby the sender tells the truth and the receiver believes the message sent by the sender, would have the highest fitness. Therefore, in this case we would expect a quick convergence to $(T,B)$. Fig. 2, panel C confirms this intuition. The case of $s>0$ is more nuanced. A theoretical analysis suggests that the system undergoes a shift in the optimal strategy profile from $(L,B)$ to $(T,N)$. Indeed, since believers have an initial advantage and since now lying is beneficial to the sender, $(L,B)$ will tend to proliferate at the initial stages of the evolution. However, as $L$ evolves, it is no longer beneficial for receivers to believe the message sent by the sender and therefore, they would start playing $N$. Since when receivers do not believe the message sent by the sender, the difference in the sender's payoff when the sender plays $T$ compared to when s/he plays $L$ is very little, this would lead the evolution to stabilize around the strategy profile $(T,N)$, with a small residual of the other strategy profiles. This nuanced evolution is reported in Fig. S2 in the SI \cite{SI}, which highlights that we do indeed see the anticipated initial increase in $(L,B)$, however, soon after, $(T,N)$ takes over. It is interesting to observe, at this stage, that in this regime where it is extremely beneficial for an agent to not be deceived but others be deceived. Reminiscent of the \emph{tragedy of the commons} \cite{milinski2002reputation,hardin2009tragedy,ostrom2002drama}, the ultimate fate here is unfavorable to all as no one earns the much coveted group payoff even for the slightest temptations to lie ($s>0$). Even the sender's payoff for lying, which instigates the transition from $(T,B)$ to $(T,N)$, is not enjoyed in the end. See panel C of Fig.~2, for a heatmap of stationary frequencies of the four strategies for $k_R=10$ and $k_W=0$, for various values of $s$ and $r$.

When $k_R=0.25$, it is still beneficial for receivers not to be deceived whenever other receivers are deceived. However, since both group interactions parameter, $k_R$ and $k_W$, are small, this time we see an evolution that more closely portrait the evolution in the pairwise game, apart from a shift to the right upon introduction of $k_R$ as opposed to the effect of $k_W$ making $(T,B)$ more prevalent for $s<0$ and promoting not-believing ($N$) for $s>0$. We report this result in panel D of Fig.~2.

To conclude this section, where we isolated the effects of $k_R$ and $k_W$, we now mention some of our results for the cases when they both act together. For $k_R=k_W$, the group payoffs of each round act like a constant addition of payoff to each strategy in the simple pairwise game, and hence, we expect that the evolution of strategies in this case resembles the pairwise sender-receiver game. We confirm our expectation in Fig. S1 in the Supplementary Information \cite{SI}. We also report the cases $(k_R,k_W)=(-0.25,0)$ and $(0,-0.25)$ in the SI. They are all very similar to the cases just reported.

\subsection{Hyperring}\label{suse:hypergraphs}

Next, we conduct simulations on a hyperring consisting of 500 nodes. The hyperring can be thought of as a lattice-equivalent for hypergraphs because of its uniform structure (see Figure 1). Apart from the total number of nodes, only one parameter has to be predetermined to uniquely define a hyperring, the size of each hyperedge. We start by analysing the case in which hyperedges have size 4. At the end of this subsection, we will discuss the dependence on the size of the hyperedge.

As a first step, we simulate the evolution for $(k_R,k_W)=(0.25,0)$ and for $(k_R,k_W)=(0,0.25)$, so that we can make a direct comparison with the results presented in the previous subsection, regarding well-mixed populations. In Fig.~3, we report the evolution for $(k_R,k_W)=(0.25,0)$ (panel A) and $(k_R,k_W)=(0,0.25)$ (panel B). In both cases, we notice that the evolution is far more nuanced on the hyperring, compared to the well-mixed populations. In particular, we notice that the ``moral strategy profile'' $(T,B)$, whereby the sender sends a truthful message and the receiver believes the message sent by the sender, which, on well-mixed populations evolve only in the trivial case of spiteful lies ($r,s<0$), in the hyperring evolves with non-zero frequency also in the domain of black lies ($r<0$, $s>0$) and in the domain of altruistic white lies ($r>0$, $s<0$). On the other hand, in both cases, the evolution of the ``immoral strategy profile'' $(L,N)$ is disfavoured, compared to the well-mixed case. We note that a similar enhancement of moral strategies has already been observed in the canonical (pairwise) sender-receiver game played on networks, compared to the case of well-mixed populations \cite{capraro2020lying}. However, we point out that higher order effects have a more nuanced effect on the evolutionary dynamics -- the evolution of the strategy $(T,B)$ on pairwise networks can be further enhanced, or completely hindered, depending on the nature of group interactions, i.e., the value of the parameters $k_R$ and $k_W$ (See Fig.~S5 in the \cite{SI}).

The bottom two figures of Fig.~3 provide a comparison between the evolution on the hyperring with that on the well-mixed populations for $(k_R,k_W)=(0.25,-0.25)$. The heatmap in panel C is for the well-mixed population, and panel D for hyperring with 4 nodes in each hyperedge. We show that in this case too, the spatial structure provided by the hyperring favours the evolution of the moral strategy profile $(T,B)$ and disfavours the evolution of the immoral strategy profile $(L,N)$, both in the domain of altruistic white lies and in the domain of selfish black lies. We point out that the two evolutions are fundamentally different. While the moral strategy profile $(T,B)$ evolves with almost zero frequency in well-mixed populations, it evolves with almost 70\% frequency in the hyperring. On the other hand, the immoral strategy profile evolves to a frequency close to 30\%.

\begin{figure}
    \centering
    \includegraphics[scale=0.34]{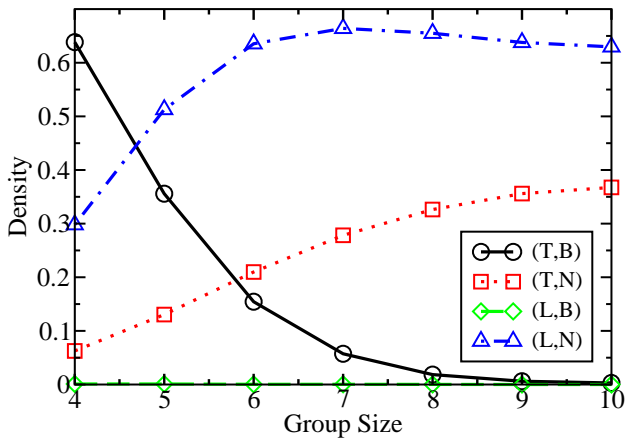}
    \caption{Stationary frequencies of the four strategy profiles for parameters $s=0.2,r=-0.3, k_R=0.25$ and $k_W=-0.25$ as a function of number of elements in a hyperedge. We show that the emergence of the moral profile due to the spatial correlations provided by the hypergraph structure is only significant when group sizes are small, and the effect dies out as we increase the group size -- the relative density of $(T,B)$ in the figure is the highest when each group contains $4$ elements, and monotonically decreases as the group sizes increase. }
\end{figure}

In the supplementary information \cite{SI}, we also report the results of the simulations for $(k_R,k_W)=(-0.25,0), (0,-0.25), (-0.25,0.25)$ and $(0.25,0.25)$. The  emergence  of  the  moral  strategy  profile, $(T, B)$, in regions of $r<0, s>0, s<-r$ (black lies) and $r>0, s<0, r<-s$ (altruistic white lie) is an especially interesting feature and can be observed in all cases, albeit with varying frequencies. Evidently, the evolution of $(T,B)$ in the regime of black lies is enhanced by a positive value of $k_R$ and negative value of $k_W$, corroborated by the high frequency of the moral profile for $(k_R,k_W)=(0.25,-0.25)$ and significantly less frequency for parameter values $(k_R,k_W)=(-0.25,0.25)$. Our results convincingly point towards the conclusion that spatial correlations provide a route for the emergence of honest signalling in groups, at least when the population consists of groups of small size.

We now analyse the dependence on the size of the hyperedges. Figure 4 reports the stationary frequencies of the four strategy profiles for parameters $s=0.2,r=-0.3, k_R=0.25$ and $k_W=-0.25$ as a function of number of elements in a hyperedge. We see that the group size is unfavourable to the evolution of moral behaviour, as the stationary frequency of the moral strategy profile $(T,B)$ decreases as a function of the group size, and becomes nearly zero for group size greater than or equal to 10. This is, a posteriori, not surprising, as, when the group size increases, the hyperring converges to a well-mixed population, which we already know to be unfavourable to the evolution of $(T,B)$ (see Section \ref{suse:well-mixed}). In the SI \cite{SI}, we also report the final densities for several other values of $k_R$, $k_W$, $r$, and $s$ (with $r,s$ either in the domain of black lies or in the domain of altruistic lies). In all cases, we found that the final density of $(T,B)$ decreases with the group size. In the domain of spiteful and Pareto white lies, clearly (T,B) either always evolve with 100\% frequency (spiteful lies) or with 0\% (Pareto white lies). This case is trivial and thus excluded from the numerical analysis.

\subsection{Real-world hypergraphs}\label{suse:real world}

The key result of the previous section is that the hyperring structure promotes the evolution of the moral strategy profile $(T,B)$, at least when the group size of the hyperedges is relatively small. One might wonder whether this is a particular feature of hyperrings with a small group size, or it holds also for other hypergraphs and, in particular, for more heterogenous hypergraphs, like the ones created from real-world data.

To answer this question, we created four hypergraphs describing real-world interactions using the publicly available SocioPatterns dataset. SocioPatterns is ``an interdisciplinary research collaboration formed in 2008 that adopts a data-driven methodology to study social dynamics and human activity. Since 2008, [they] have collected longitudinal data on the physical proximity and face-to-face contacts of individuals in numerous real-world environments''\footnote{http://www.sociopatterns.org}. In particular, the SocioPatterns datasets record face-to-face interactions with a temporal resolution of 20 seconds. This allows us to check whether individuals are truly interacting as a group, and this allows us to create a hypergraph as follows. For every 20 second window, we create a network of interactions and catalogue all the maximal cliques beyond size 2. Repeated appearance of clique, beyond a specified threshold, is treated as a group interaction and we create a hyperedge consisting of the nodes in the recurring clique. A similar approach of creating hypergraphs has already been adopted to study opinion dynamics on hypergraphs \cite{sahasrabuddhe_modelling_2020}, as well as studying the temporal nature of higher order interactions \cite{cencetti_temporal_2020}.

\begin{figure*}
    \centering
    \includegraphics[scale=0.53]{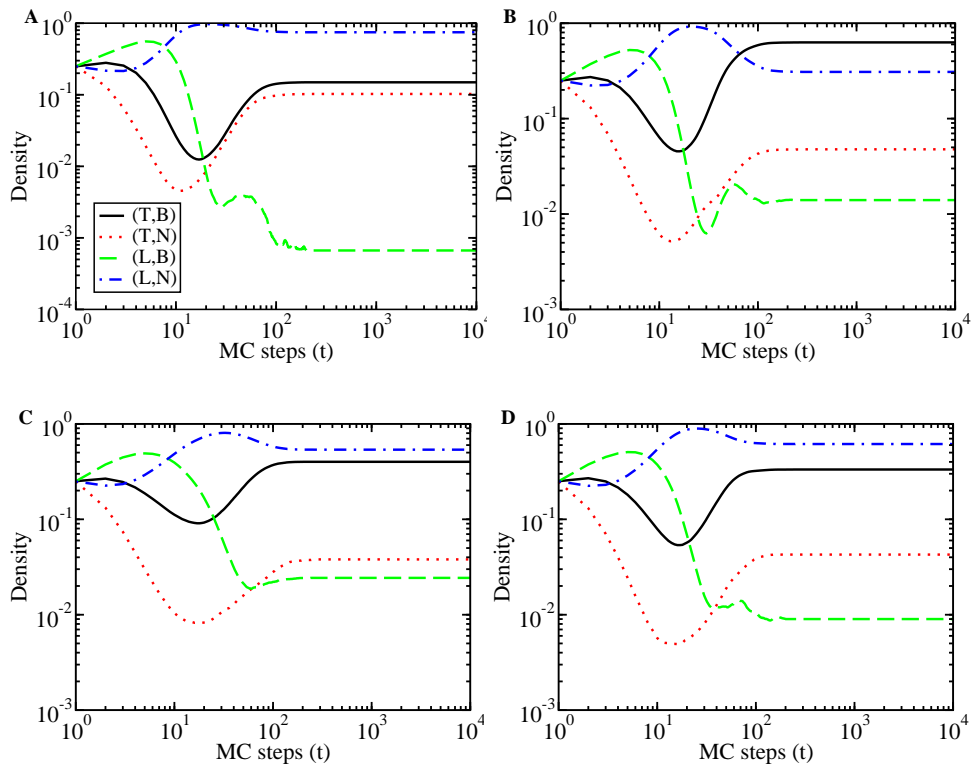}
    \caption{Persistence of honest signalling and believing behaviour $(T,B)$, demonstrated on four hypergraphs generated using real world interaction patterns, namely primary school (panel A), conference (panel B), high school (panel C), and workplace (panel D) datasets (descriptions in text) for $k_R=0.25, k_W=-0.25$, $s=0.2$ and $r=-0.3$. These simulations provide evidence for our observation on hyperrings than higher order interactions, coupled with spatial correlations allow for the emergence of the moral profile $(T,B)$.}
\end{figure*}

In doing so, we created four hypergraphs starting from the following datasets that we downloaded from SocioPatterns:
\begin{enumerate}
    \item Primary Schools dataset, formed by school children and teachers at a primary school in France.
    \item Conference dataset, formed by participants at the 2009 SFHH conference in Nice, France.
    \item High school dataset, formed by students at a high school in Marseilles.
    \item Workplace dataset, formed by the staff at an office building in France
\end{enumerate}

These datasets represent a diverse variety of social situations and thus, provide the ideal setting for our simulations. Figure 5 reports the time evolution of the four strategy profiles for each of these four real-world hypergraphs. We note that, in each hypergraph, the moral strategy profile $(T,B)$ persists with a non-zero frequency. In the primary school hypergraph (panel A), it evolves with frequency close to 17\%; in the high school hypergraph (panel B) it evolves with frequency around 30\%; in the workplace hypergraph (panel C), it evolves with frequency around 32\%; and in the conference hypergraph (panel D), it even evolves with frequency above 60\% \footnote{We wondered whether these differences were driven by underlying differences in the group sizes of the hyperedges, along the lines discussed at the end of the previous subsection. With this in mind, we plotted the group size distribution for all the graphs but could not find a systematic structural feature for the differences in evolution of (T,B).}.

\section{Discussion}

In our work, we provided a natural extension to the sender-receiver game which allowed us to study the impact of group interactions on the evolution of honesty. Different combinations of parameter values in our model correspond to different types of higher order effects in strategic interactions among agents. Some particularly relevant choices of parameters, from a practical perspective, include:
    \begin{enumerate}
       \item \emph{$r<0$ and $0>k_R>k_W$}. It is beneficial for each receiver if they do not get deceived and also the other receivers in the group do not get deceived.\\
       \emph{Example.} During the COVID-19 crisis, there have been several incidents caused by the spreading of fake news. For example, in India a rumour spread about some trains being arranged by the government for the labourers that were stuck away from home. This caused a gathering of thousands of people at a major railway station \footnote{A brief report regarding this incident that occurred in Mumbai, India, can be found at: https://indianexpress.com/article/cities/mumbai/mumbai-news-report-misleading-led-to-gathering-at-bandra-says-court-6370029/}. We can formalize this situation with: $k_R, k_W<0$, because the more receivers get deceived, the more social distancing rules are violated, leading to an increase in the COVID-19 transmission, with potentially negative consequences also for receivers that were not deceived; $k_R>k_W$, because, everything else being equal, it is more likely to get infected if being deceived (i.e., gathering at the railway station), than if not; $r<0$, because receivers who go to the railway station spend money and time for no reason.
       \item \emph{$r<0$, $k_R>k_W$, and $k_R>0 $}. It is beneficial for each receiver if they do not get deceived, but other receivers in the group get deceived.\\
       \emph{Example.} At the beginning of the COVID-19 crisis, there has been a shortage of essential goods, including groceries and household items of daily use. Suppose there is a rumour spreading that these goods are available at shop $X$ and not at shop $Y$ (where they are actually available). Then, it is in the receiver's best interest if they do not get deceived but other people do, in order to minimize the competition for the utilities in shop $Y$.
       \item \emph{$r>0$ and $0>k_W>k_R$.} It is beneficial for each receiver if they get deceived but the other receivers in the group do not get deceived.\\
       \emph{Example.} Any white lie being told in a population of competing receivers belongs to this case. For instance, suppose that two sport teams are competing for an important match and that, right before the match, the presidents of the two teams are informed about some tragic event that might lower the performance of the teams, for example, the sudden death of someone well known in their field. The presidents have to decide whether to tell the truth to the coaches and the teams before the match or not. In this case, for each team, it is better to believe the white lie, while the other team does not. Indeed, in this case, it is more likely to win the game, because the other team lowers its performance.
       \item \emph{$r>0, k_W>k_R, k_W>0$}. It is beneficial for each receiver if they get deceived and that also the other receivers in the group get deceived.\\
       \emph{Example.} This is a very general case, because there are a number of examples where cooperative behavior, that benefits everyone, emerges from collective beliefs which are not supported by evidence. The examples include various cultural beliefs.
   \end{enumerate}

By performing Monte Carlo simulations of the proposed model, we first explored in detail the stationary frequencies of the different strategy profiles in well-mixed populations across the parameter space of the model. In doing so, several novel features emerged out of group interactions, which were not observed in the canonical pairwise sender-receiver game. Also by isolating the group effects, we gained further insight into how the evolution is modified upon the inclusion of group interactions in well mixed populations.

But, of course, real populations are often structured, and, in many cases, individuals do not interact in pairs, but in groups. These higher-order structures are thus important for a realistic modelling of any complex system. To probe them, we used the formal paradigm of hypergraphs, which allowed us to model networked systems with group interactions. We first performed simulations on hyperrings, which, owing to its uniform structure, can be thought of as the hypergraph version of a lattice. We observe that the evolution of strategies changes drastically, and, in particular, we identified the regions in the parameter space where the `moral strategy profile', whereby senders tell the truth and receivers believe the message sent by the senders, evolve, which were not observed in well mixed populations. In particular, we find that new values of the parameters emerge, which allows agents to overcome the temptation to lie ($s>0$), even for comparatively small values of $k_R$ and $k_W$. We further established that this effect, which was not present in well-mixed populations, is strongest when the group size is small and decays very quickly with the increasing sizes of groups. This decrease in morality as a function of the group size is not surprising, as our model of the hyperring converges to well-mixed populations as the size of the hyperedges increases. What is surprising is the fact that the moral strategy profile does evolve in small groups. Moreover, this is not just a mathematical curiosity due to the special hyperring structure, as we found a qualitatively similar result in all real-world hypergraphs that we have built from the SocioPatterns dataset: the strategy profile $(T,B)$ persists in all real-world hypergraphs that we have built, although with different final densities.

Recent works have explored the evolution of honesty on only the canonical, pairwise sender-receiver game in well-mixed populations and on networks \cite{capraro2019evolution,capraro2020lying}, as part of a research direction to study the evolution of \emph{moral behaviour} \cite{capraro2018grand,kumar2020evolution}. This work constitutes the first systematic study of group effects in signalling games. As any first study, also this has some limitations that can suggest directions for future work. For example, we considered only the case in which there is one sender and multiple receivers. However, in reality, sometimes there are multiple senders that compete among themselves, such as news outlets which compete for who has more receivers. Future work could extend our formalization to include competition among senders. Also, as real higher-order networks, we considered only four hypergraphs downloaded from the SocioPatterns database, which refer to physical interactions in a primary school, a high school, a conference, and an office. It is possible that these interactions are spatially different from virtual interactions that happen online. Understanding the effects of the network structure on the spread of misinformation online is certainly an important direction for future work.

This study also contributes to our understanding of the effect of group size on the evolution of morality. Previous work has mainly focused on another form of morality, cooperative behavior \cite{capraro2018right}. In this context, a line of work using several techniques, ranging from numerical simulations, to mathematical analyses and behavioral experiments, has found that the relationship between group size and cooperation is very nuanced and much depends on the particular experimental paradigm being used to formalize cooperative behavior \cite{olson2009logic,komorita1982cooperative,grujic2012three,nosenzo2015cooperation,mcguire1974group,isaac1994group,zhang2011group,esteban2001collective,oliver1988paradox,chamberlin1974provision,barcelo2015group,capraro2015group}.

Cooperation, one of the most well studied topics in the field of evolutionary game theory \cite{santos2008social,ohtsuki2006simple,santos2005scale,santos2006cooperation,perc2017statistical,burgio2020evolution}, is only one specific type of moral behaviour among several others \cite{capraro2018right}, among which honesty is also present \cite{biziou2015does}. However, unlike cooperation, other moral behaviors have not received significant theoretical attention, and certainly not from the point of view of group interactions. In a society as complex as ours, it is expected that the phenomena that emerge out of it cannot be explained solely through pairwise interactions between individuals as many interactions naturally happen in groups. We hope that our work provides a stepping stone towards bridging this gap and encourages further studies devoted to group interactions. An interesting future direction also emerges on the experimental side, where behavioral changes in strategic interactions between individuals can be explored, in situations where their rewards are entangled with each other.

\section*{Acknowledgments}
A.K. acknowledges support from  the Prime Minister's Research Fellowship of the Government of India. M.P. acknowledges funding from the Slovenian Research Agency (Grant Nos. P1-0403, J1-2457, and J1-9112).

\providecommand{\noopsort}[1]{}\providecommand{\singleletter}[1]{#1}%

 \clearpage

\onecolumngrid

\setcounter{page}{1}
\renewcommand{\thepage}{S\arabic{page}}
\setcounter{equation}{0}
\renewcommand{\theequation}{S\arabic{equation}}
\setcounter{figure}{0}
\renewcommand{\thefigure}{S\arabic{figure}}
\setcounter{section}{0}
\renewcommand{\thesection}{S\arabic{section}}
\setcounter{table}{0}
\renewcommand{\thetable}{S\arabic{table}}

\maketitle

\onecolumngrid


\begin{center}
\large\textbf{Supplementary results}
\vspace{10mm}
\end{center}

In what follows, we provide additional figures that support the results and arguments presented in the main paper.

\begin{figure}[h]
    \centering
    \includegraphics[scale=0.74]{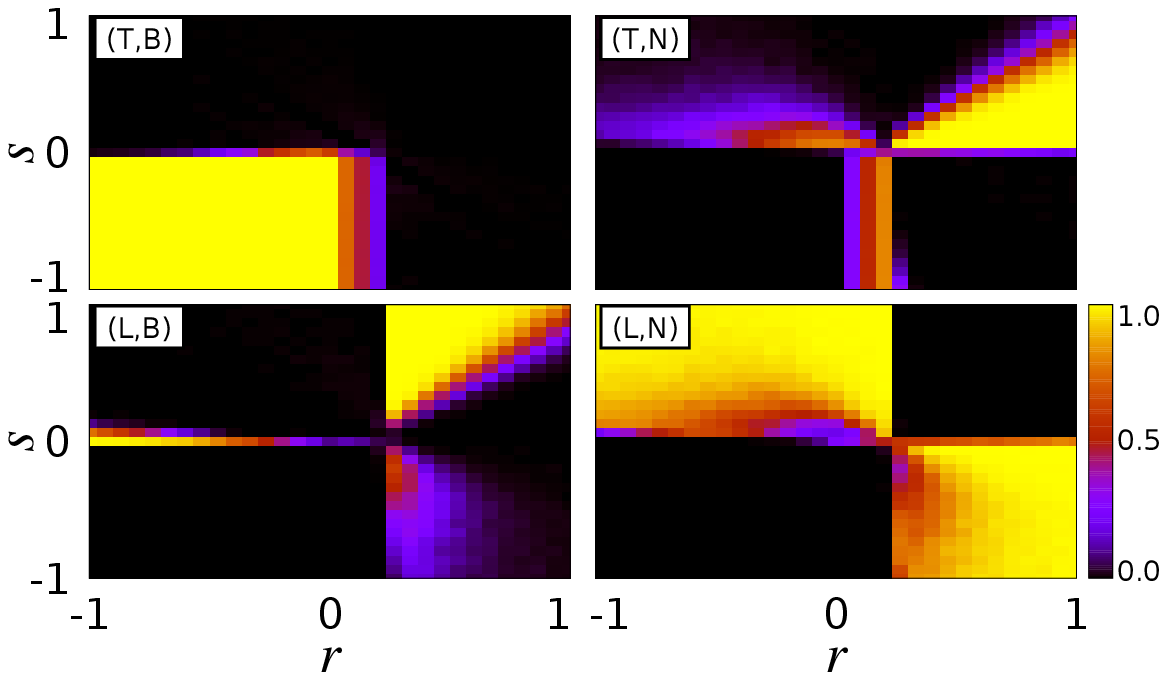}
    \caption{Final densities of the four strategy profiles $(T,B)$, $(T,N)$, $(L,B)$ and $(L,N)$, when $k_R=0.25$ and $k_W=0.25$.}
    \label{fig:my_label}
\end{figure}

\begin{figure}
    \centering
    \includegraphics[scale=0.4]{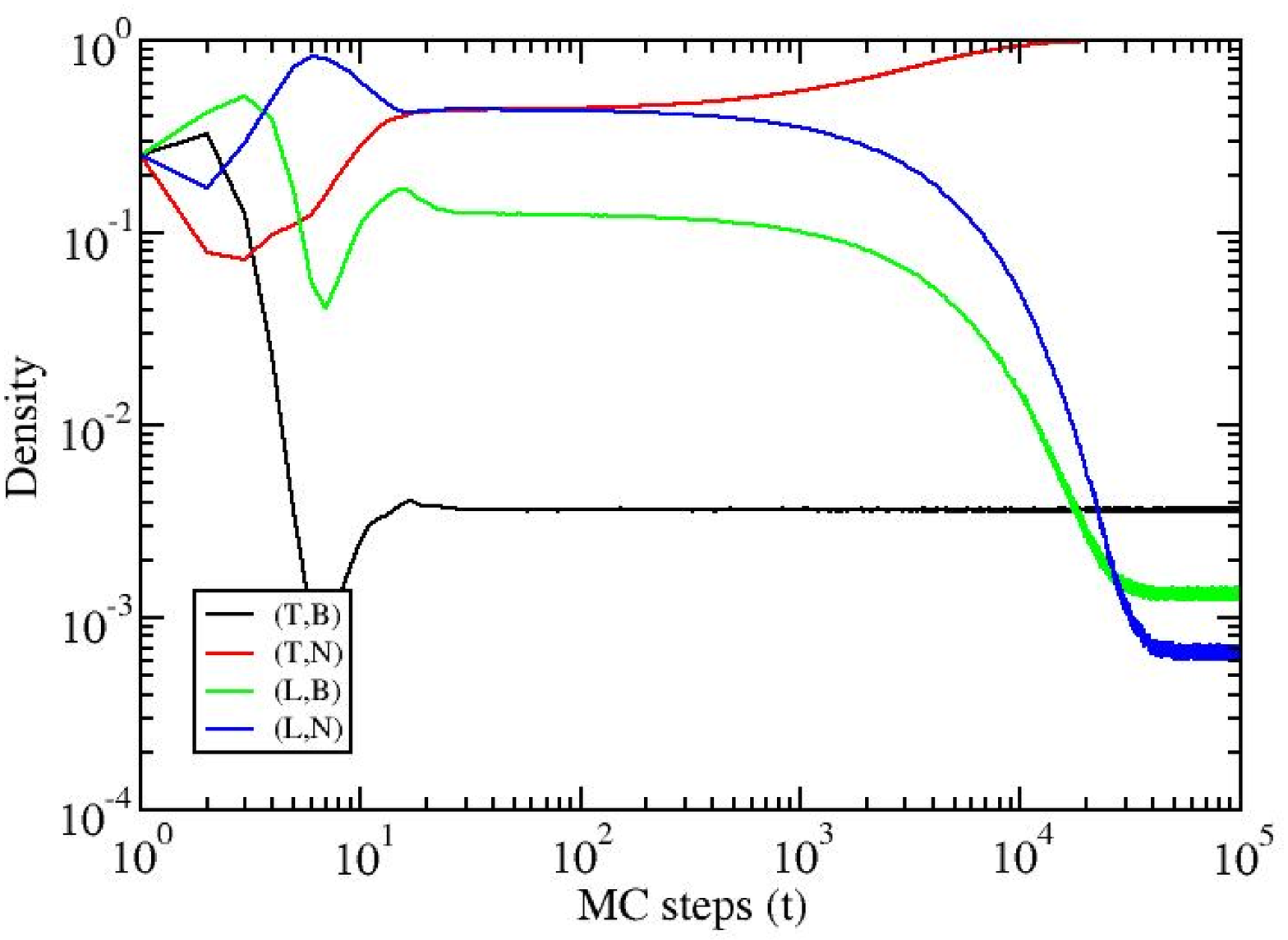}
    \caption{The time evolution of strategies for $s=0.5, r=0.5$, $k_R=10$ and $k_W=0$ in well mixed populations, depicting the slow relaxation to equilibrium.}
    \label{fig:my_label}
\end{figure}

\begin{figure*}
{\epsfig{file=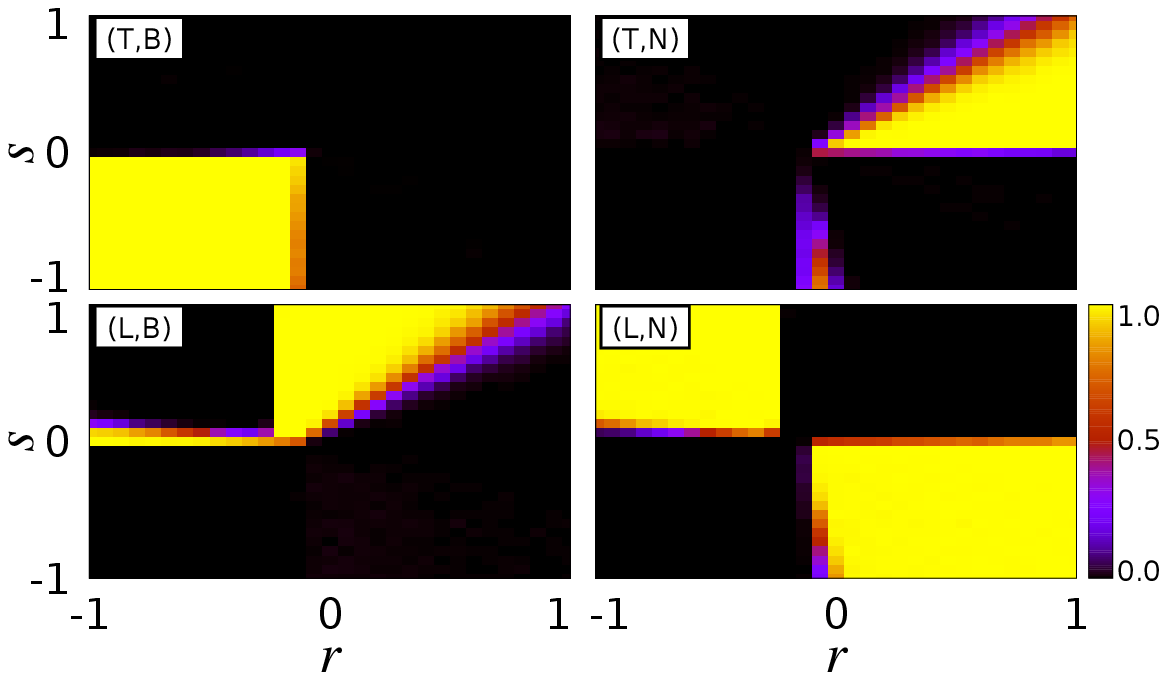,width=8.5cm}}
{\epsfig{file=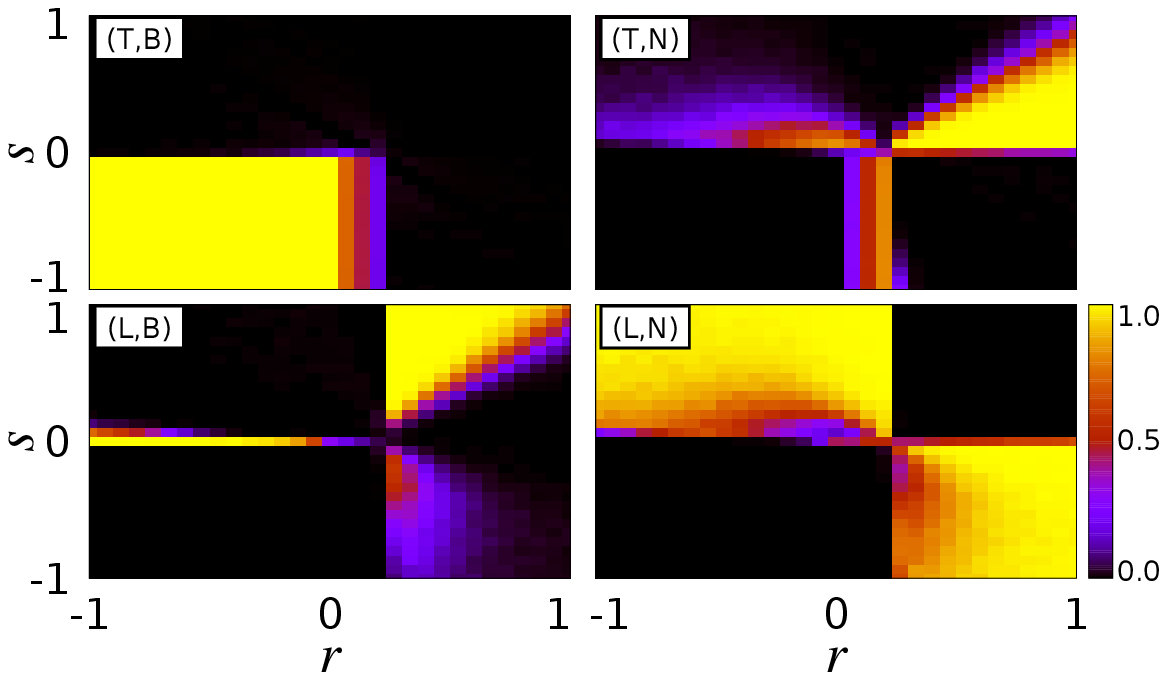,width=8.5cm}}
\caption{Stationary frequencies of the four strategies on the hyperring with 4 nodes in each hyperredge: (Left) $(k_R,k_W)=(-0.25,0)$  and (Right) $(k_R,k_W)=(0,-0.25)$ .}
\end{figure*}

\begin{figure*}
{\epsfig{file=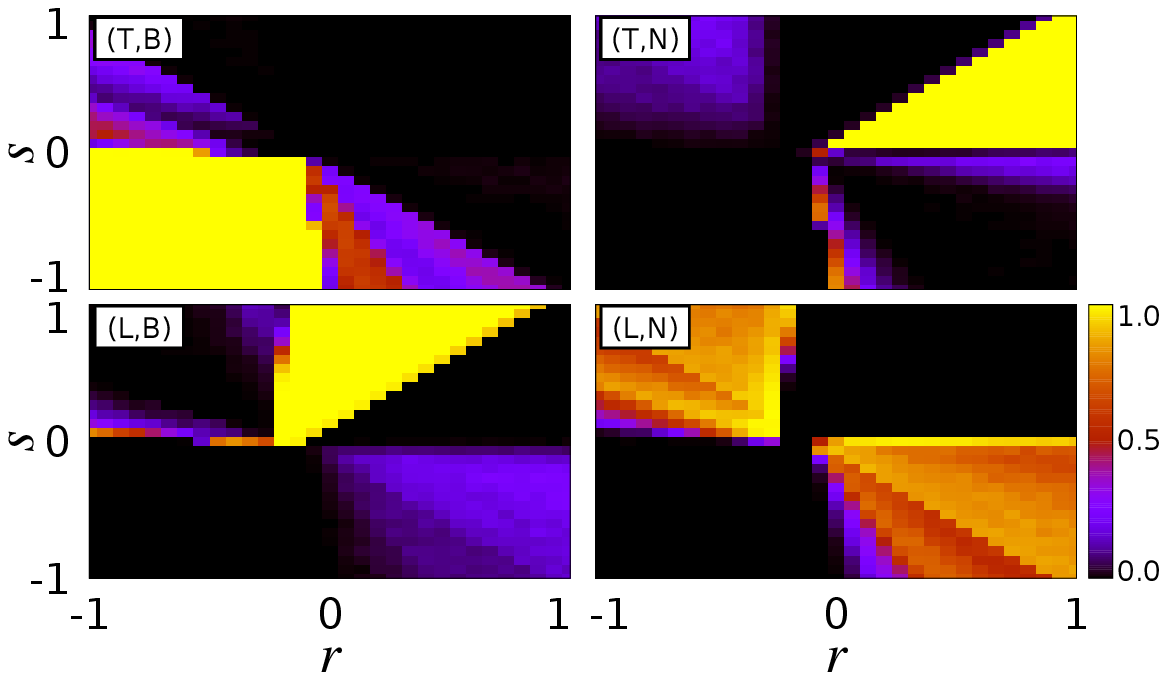,width=8.5cm}}
{\epsfig{file=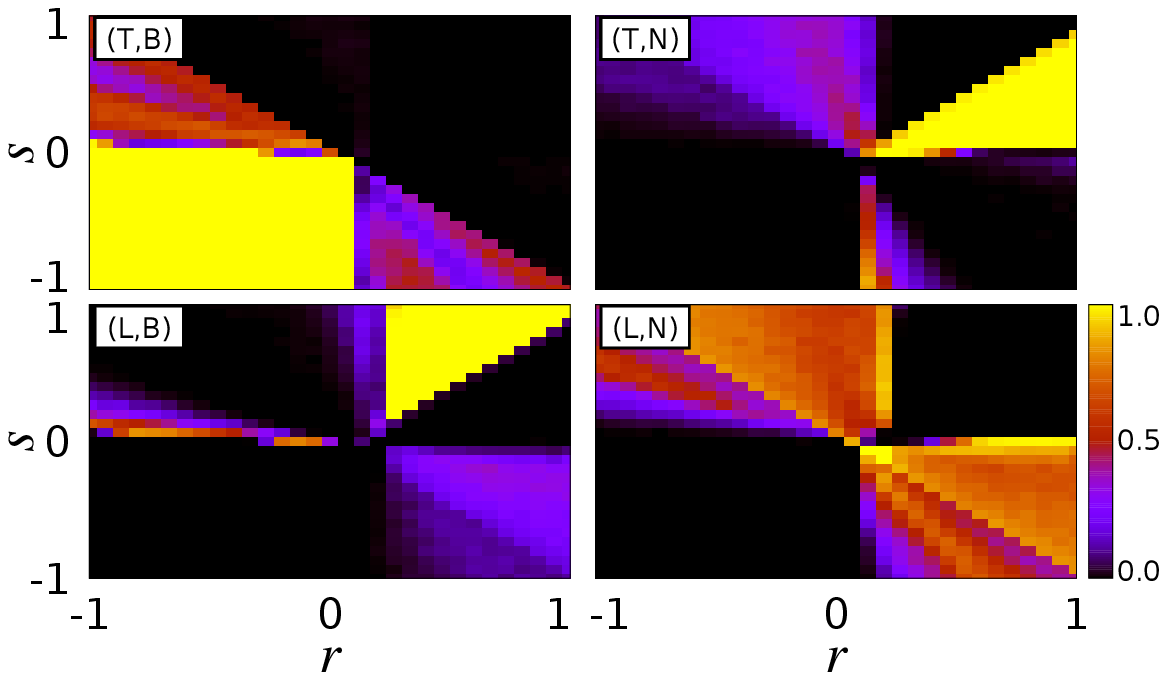,width=8.5cm}}
{\epsfig{file=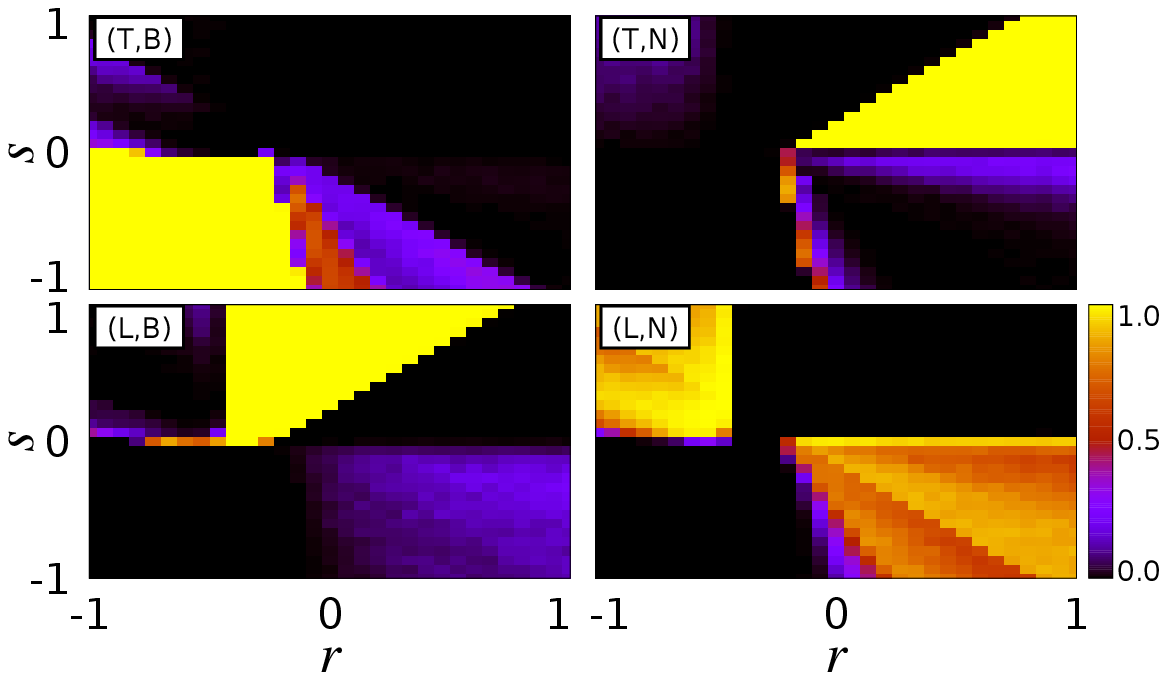,width=8.5cm}}
{\epsfig{file=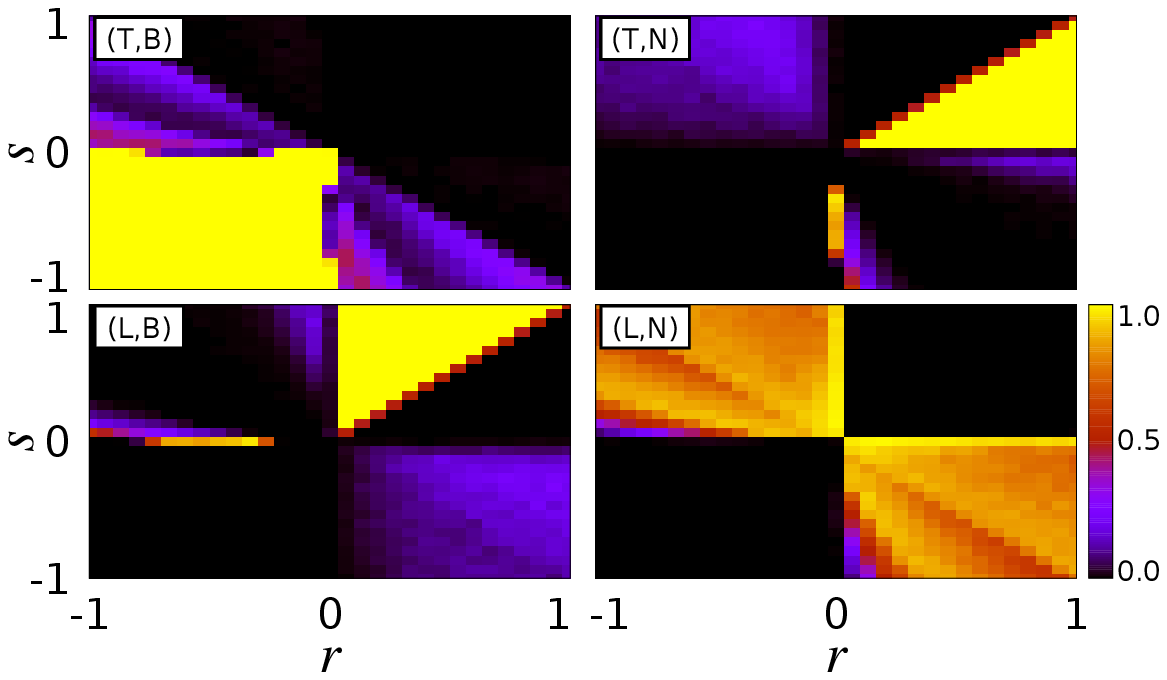,width=8.5cm}}
\caption{Stationary frequencies of the four strategies on the hyperring with 4 nodes in each hyperredge: (Top left) $k_R=-0.25,k_W=0$, (Top right) $k_R=0,k_W=-0.25$, (Bottom left) $k_R=-0.25,k_W=0.25$, (Bottom right) $k_R=0.25,k_W=0.25$.}
\end{figure*}

\begin{figure*}
{\epsfig{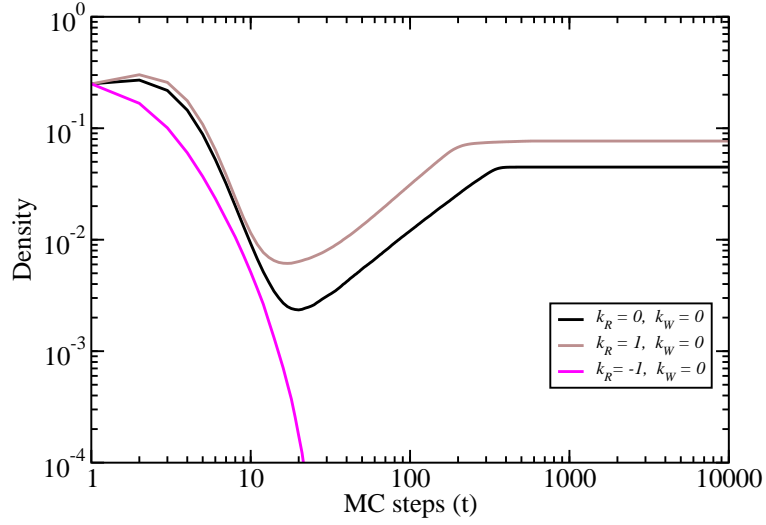}}
\caption{Evolution of the moral strategy $(T,B)$ for ($k_R$,$k_W$)$=(0,0), (1,0),$ and $(-1,0)$. The parameter values $s=0.2$ and $r=-0.3$ are fixed for all three curves. The ($k_R$,$k_W$)$=(0,0)$ curve depicts the simulation on the projected pairwise network of the hypperring (where each hyperredge is replaced by a clique with only pairwise links between each member of the hyperedge). While it was known that networks facilitate the evolution of moral behavior in the pairwise sender-receiver game, it is clear that group interactions can further promote the strategy $(T,B)$ (brown curve) or even hinder it (pink curve), depending on the nature of group interactions (the value of $k_R$ and $k_W$).}
\end{figure*}

\begin{figure*}
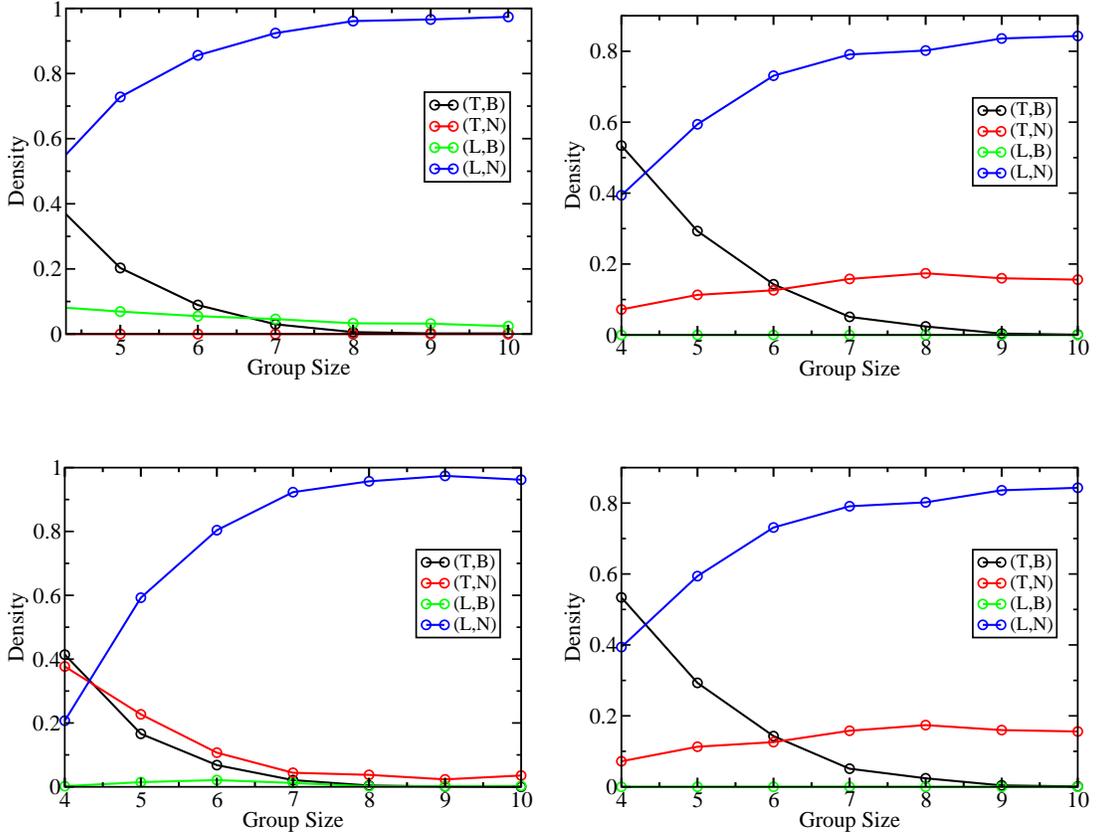

\centerline{\epsfig{file=sfig6a.eps,width=7cm}\hspace{4mm}\epsfig{file=sfig6b.eps,width=7cm}}
\vspace{10mm}
\centerline{\epsfig{file=sfig6c.eps,width=7cm}\hspace{4mm}\epsfig{file=sfig6d.eps,width=7cm}}
\caption{Stationary frequencies of the four strategy profiles in the hyperring, as a function of number of elements in a hyperedge, for parameters (Top left) $s=-1,r=1, k_R=0.25$ and $k_W=0$, (Top right) $s=0.2,r=-0.3, k_R=0$ and $k_W=-0.25$, (Bottom left) $s=-0.8,r=0.2, k_R=0$ and $k_W=-0.25$, and (Bottom right) $s=0.1,r=-0.8, k_R=0$ and $k_W=-0.25$.}
\end{figure*}

\end{document}